# Experimental Demonstration of Secure Frequency Hopping Communication Enabled by Quantum Key Distribution

Bernardo A. Huberman, Bob Lund, Jing Wang, and Lin Cheng

Next-Gen Systems Group, CableLabs, Louisville, Colorado 80027, USA

*Abstract*— We propose and experimentally demonstrate a method of frequency hopping spread spectrum communication using a quantum key distribution network to deliver the frequency hopping pattern for secure wireless communications. Results show low interception and jamming probabilities.

## I. Introduction

Frequency hopping spread spectrum (FHSS) is a robust and efficient modulation technique that has revolutionized the field of wireless communications. With the demand for reliable and secure wireless networks, FHSS has emerged as a promising solution, providing enhanced resistance against interference and eavesdropping. This technique involves rapidly switching carrier frequencies within a designated frequency band, thereby distributing the transmitted data over a wide range of frequencies. By employing this dynamic frequency allocation scheme, FHSS mitigates the impact of narrowband interference and improves the overall performance of wireless systems.

While FHSS offers numerous benefits, it also faces certain challenges that need to be addressed for optimal performance. As the FHSS keeps changing central frequencies, it is extremely challenging for the receiver to synchronize to the next frequency accurately if without coordination [1]. Therefore, the FHSS sender and potential receiver need to agree on the order of frequencies and intervals to be used during the communication. This is done by sharing a cryptographically encrypted sequence of frequencies or utilizing an agreed-upon hopping algorithm for selecting a frequency for each interval. However, either method is susceptible. The cryptographic method is only as secure as the crypto algorithm used. The hopping pattern can be predicted by advanced machine learning and, in certain scenarios such as military applications or environments, sophisticated adversarial signal analysis. This becomes even more tenuous with the advent of quantum computers. Consequently, addressing prediction-related challenges is crucial to maintaining the integrity and confidentiality of FHSS-based systems.

Quantum key distribution (QKD) provides information-theoretic security and truly random key bits [2]. The security of quantum key bits is guaranteed by quantum mechanics. An unknown quantum bit (qubit) cannot be copied due to the no-cloning theorem and can neither be measured without intervention since observation causes perturbation [3, 4]. Since most qubits are carried by photons, so far as we know, most QKD systems are used to protect optical communications and wireless communication has not benefited from the information-theoretic security offered by QKD yet. In this paper, we leverage the security provided by an optical-fiber QKD network to enable a more secure and robust FHSS wireless communication, compared with existing deterministic algorithms, in light of the growing capabilities of machine learning and quantum computing-based eavesdropping and jamming.

## II. FHSS with Quantum Key Distributed Hopping Pattern

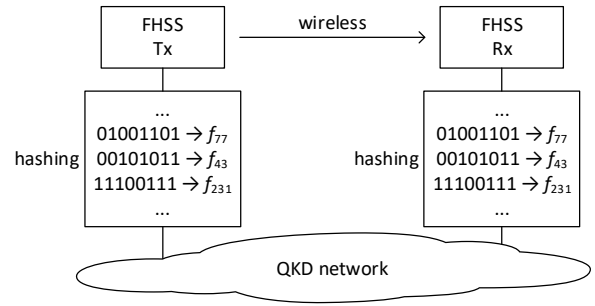

Fig. 1. Operation principles of FHSS communication enabled by QKD networks for frequency hopping pattern exchange.

Fig. 1 shows the operation principles of FHSS communication enabled by a QKD network for frequency hopping pattern exchange. In a QKD network, nodes are interconnected by optical fibers and exchange quantum keys in the physical layer, which can be used to encrypt network traffic in all the above layers. It should be noted that this architecture is protocol-agnostic and any existing QKD systems/protocols can be used for this purpose.

The FHSS transmitter and receiver go through a protocol to find a common quantum key ID that each side uses to select a common quantum key. The QKD sequence is used to generate the hopping frequency pattern. In the example of Fig. 1, there are up to 256 hopping frequencies, as the QKD sequences are partitioned into 8-bit bytes. Each byte as a frequency index is interpreted as a frequency value through a hash table. The sender and receiver agree on how much time is spent at each frequency, i.e. the hopping time interval.

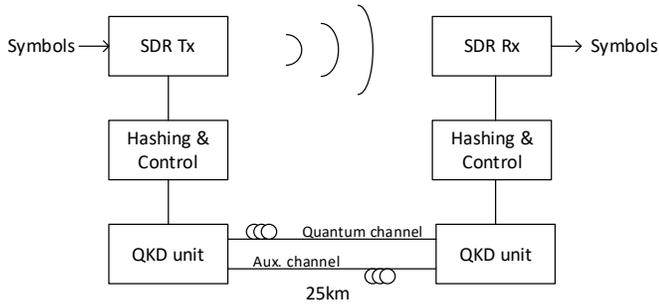

Fig. 2. Experimental setup of proposed system using SDR for wireless

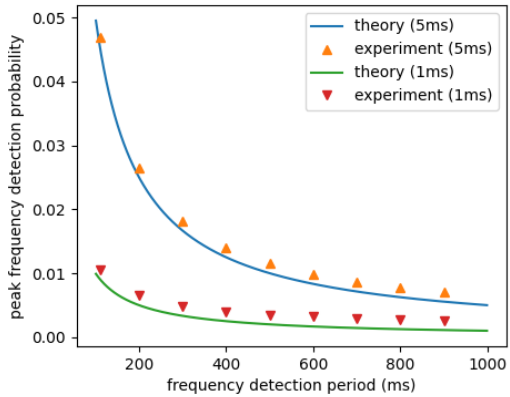

Fig. 3. Peak frequency detection probability vs. frequency detection period under different hopping time intervals

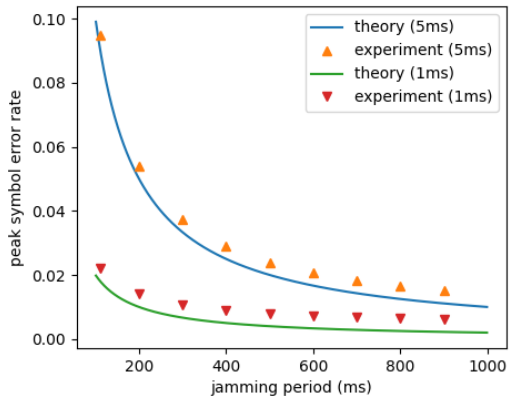

Fig. 4. Peak symbol error rate vs. jamming period under different hopping time intervals (without FEC)

## III. Experiment and Results

Fig. 2 shows the experimental setup of the proposed system. We used a Clavis3 QKD platform developed by ID Quantique [5]. The two QKD units are linked by two 25-km optical fibers, one for the QKD link, and the other for auxiliary classical communications, such as synchronization, key sifting, and post-processing. It is based on the coherent one-way (COW) protocol, where an intensity modulator encodes qubits into time bins [6]. The 0-bit, 1-bit, and decoy states are represented by patterns with an optical pulse in the first, second, or both time bins, respectively. The pulses are then attenuated to the single-photon level before transmission. After raw key exchange, the raw keys are post-processed to correct errors and reduce the information an eavesdropper has to an arbitrarily low level. In Clavis3, key sifting, error correction, and privacy amplification are implemented and automated by FPGA to allow a secure key exchange. Our system has a quantum bit error rate (QBER) of around 3.5% and a key rate of 2000 b/s. The Clavis3 system also has a key management interface to transfer the keys from the QKD system to our frequency-hopping system over the secured ETSI REST API.

The FHSS wireless system is implemented on two software-defined radio (SDR) transceivers. The hash table interprets the QKD information into 128 discrete frequency values which drive the center frequency of an LTE-like signal that has a half- and sub-frame duration of 5 ms and 1 ms, respectively. These durations determine the hopping time intervals that we have chosen. To examine the robustness against frequency detection (eavesdropping), we tested peak detection probability with respect to different frequency detection periods under the two hopping time intervals. Fig. 3 shows that the proposed system has a very small penalty compared with the theoretical ideal probability. This is mainly due to the limited bandwidth of the SDR system We then examined the robustness against signal jamming. We tested the peak symbol error rate with respect to different jamming periods under the two hopping time intervals. Our jamming signal is approximately 20 dB stronger than the FHSS signal. Each interval the jamming signal overlapping or being adjacent to the FHSS signal causes an undecodable symbol without FEC counted as a symbol error. Fig. 4 shows that the proposed system has a very small penalty compared with the theoretical ideal symbol error rate, partially attributed to the existence of out-of-band interference and also the limited bandwidth of the SDR system.

## IV. Conclusion

We propose and experimentally demonstrate a method of FHSS communication using a QKD network to exchange frequency hopping patterns for secure wireless communications. Experimental results show that the proposed system has a very low penalty compared with theoretical ideal probabilities in terms of interception and jamming.